\documentclass[twocolumn,twoside]{article}
\usepackage{graphics}
\newcommand{\h}{\linebreak \hspace*{3ex}}
\newcommand{\hb}{\\ \hspace*{2ex}}

\begin{document}
\title{On the possible  nature of Bp-Ap Stars: an application to HD101065 and HR465}
\author{V.F.\,Gopka$^{1}$, O.M.\,Ulyanov$^{2}$, S.M.\,Andrievsky$^{1}$\\[2mm] %English only
\begin{tabular}{l}
 $^1$ Department of Astronomy, Odessa National University\hb
 T.G. Shevchenko Park, Odessa 65014 Ukraine, {\em gopka.vera@mail.ru; scan@deneb1.odessa.ua}\\[1mm]
$^2$  Institute of Radio Astronomy of National Academy of Sciences
of Ukraine\hb 4 Chervonoprapona str., Kharkov 61002, Ukraine, {\em
oulyanov@rian.kharkov.ua}\\[2mm]
\end{tabular}
% no tabular, if one affiliation only
}
\date{}
\maketitle

{\bf Abstract}\\[1mm]

We have proposed the new explanation of some magnetic chemically
peculiar (MCP) stars anomalies, which is based on assumption that
such stars can be the close binary systems with a secondary
component being neutron star. Within this hypothesis one can
naturally explain the main anomalous features of MCP stars: first
of all, an existence of the short-lived radioactive isotopes
detected in some stars (like Przybylski's star and HR465), and
some others peculiarities (e.g. the behavior of CU Vir in radio
range, the phenomenon of the roAp stars).\\[2mm]

{\bf 1. Introduction}\\[1mm]

It is known that about 10-15\% of upper main sequence stars have
atmospheres with anomalies in elemental abundance (Bagnulo, 2002).
These stars are often classified as magnetic chemically peculiar
stars (MCP, or Bp-Ap stars) and nonmagnetic (Hg-Mn and Am-Fm)
stars.  Hg-Mn stars have a higher temperature than MCP stars and
overlap with MCP stars in the region of the higher temperatures,
while Am-Fm  stars have the lower temperatures than MCP stars.

Among the MCP stars one can select the following subclasses (Kurtz
\& Martinez, 2000):\\[1mm] 1. SrCrEu stars (spectral classes
A3-F0).\\[1mm] 2. Si stars(B8-A2).\\[1mm] 3. He-weak, Si, SrTi
stars (B3-B7).\\[1mm] 4. He-strong stars (B1-B2).
\\[2mm]

Some of the peculiarities seen in the chemically peculiar stars
can be explained by considering different non-nuclear processes
first proposed by Michaud (1970) and then elaborated in a series
of the subsequent papers (Turcotte et al., 1998; Turcotte et al.,
2000; Richard et al., 2002).\\[2mm]

{\bf 2. MCP stars: the main features}\\[1mm]

The main characteristics of MCP stars are the following:

1. MCP stars show periodic variations of the light, spectrum and magnetic
field. It is believed that these variations are due to the rotation
of the star. MCP stars has been regarded as rigid magnetic dipoles with
respect to the rotation axis (Babcock, 1949; Stibss, 1950). In the later
works it was obtained that MCP stars can be stars with a multi-pole
magnetic field (Landstreet, 1990; Mathys, 1991).

2. Some MCP stars are the source of radio emission (Drake et al.,
1987). The flat spectra and variability of radio emission on time
scales as short as a few hours, as well as the well known strong
magnetic field indicate a nontermnal gyrosynchrotron emission
process (Lynsky et al., 1989). Later Lynsky (Lynsky et al., 1992)
remarked, that all observed properties of radio emission from
these stars may be understood as optically thick gyrosynchrotron
emission from  a nonthermal distribution of electron emitters. The
radio luminosities of MCP stars are correlated with effective
temperature and magnetic field strength (Drake et al., 1994).

3. Some MCP stars are detected as X-ray sources (Drake et al.,
1994). They have been investigated during the ROSAT All-Sky
Survey. Drake (1998) emphasized that only 10 X-ray sources
detected at the positions of 100 magnetic Bp-Ap stars.

4. Shore \& Brown (1990) showed that MCP are characterized by an
anisotropic stellar wind. MCP stars have been known to have
moderate circular polarization (Drake et al., 2002). Trigilio et
al. (2000) showed that only CU Vir (Si-star) have a strong 1.4 GHz
flux enhancement around phases 0.4 and 0.8, with a right circular
polarization of almost 100\%. Leto et al., (2007) remarked that
only MCP star with a high photospheric temperature develop a
radiative-driven stellar wind, which is the cause of the radio
emission.

5. Among MCP stars there is known the phenomenon of roAp-stars.
roAp stars are the main sequence SrCrEu chemically peculiar stars
from mid-A to early-F, which pulsate with periods in the range of
5-16 min and amplitudes $>0.016$ mag (Kurtz \& Martinez, 2000).
The mechanisms responsible for an existent of the roAp star
pulsations remain unknown. HD101065 is the first Ap star where
pulsations were detected. Kurtz \& Wegner (1979) detected the
well-defined pulsations with a peak-to-peak amplitude of 0.012 mag
and a period of 12.14 min (Kurtz \& Wegner, 1979).

6. The most unusual feature is the chemical composition of MCP
stars. The first studies of the abundance anomalies in magnetic
stars $\alpha2$ CVn, HD133029, HD151199 showed that there can be
the certain nuclear reactions in the star (Burbidge et al., 1958).
Now it is accepted that detected over- and/or under-abundance of
some elements do not reflect the chemical composition of the
entire star, but only its photosphere. There are the most extremal
cases among the MCP stars. For example HD101065 with its chemical
composition is the most unusual roAp star. The star HD101065 in
fact is the unique astrophysical laboratory for understanding and
exploring the extreme phenomena of the stellar evolution. This
stars shows the high lithium abundance, as well as unusual
isotopic lithium ratio (Shavrina et all., 2000). The presence of
the short-lived lanthanide promethium $Z=61$ was noted by Wagner
\& Petford (1973), Cowley et al. (2004), Fivet et al. (2007).
The lines of some short-lived transbismuth isotopes were detected
by Gopka et al. (2004), Gopka et al. (2005), Bidelman (2005),
Quinet et al. (2007). Cowley \& Hubrig (2002), Cowley et al.
(2007) found anomalous isotopic ratio of Ca in the HD101065
atmosphere.\\[2mm]

{\bf 3. About nucleosynthesis on the stellar atmosphere}\\[1mm]

The idea about the possibility of the nucleosynthesis in the
stellar atmosphere is not new. In the mid of past century the
first papers concerning the chemical composition of Ap stars
showed that overabundance of heavy elements really exist. In some
cases overabundance of some elements can reach 6 dex and more
comparing to the solar abundance.  E.M. Burbidge and G.R. Burbidge
wrote: "The list of elements with the increased content led us to
the thought, that in this case we deal with the nuclear, not with
atomic processes and that somewhere and somehow the neutrons take
part in them." (Burbidge \& Burbidge, 1996).

Nevertheless, the mechanism responsible for such uncommon
processes was not identified at that time. Recently, Goriely
(2007) showed that nucleosynthesis of heavy (including
radioactive) elements can occur in the star's atmosphere due to
the high-energy particles entering the atmosphere of the star. At
the same time, the origin of such particles was not clearly
specified. This idea was also discussed in Arnold et al. (2007).

The r-process has been frequently mentioned in relation to the
abundance anomalies observed in Ap stars (Cowley et al., 1973). In
particular, Burbidge (1965) indicates the fact that the explosion
remnants of a more massive star can be possible explanation of
the origin of the peculiarity. For instance, a supernova explosion
in the binary system and following contamination of its companion
star with freshly synthesized material could, in principle, explain
the peculiarity, observed in HD101065. The only problem is that in
this case one has to suppose that such an event should have
happened not very long ago, since we are observing now the signs
of the short-lived isotopes in atmosphere of this star.\\[2mm]

{\bf 4. Possible origin of roAp stars HD101065 and HR465}\\[1mm]

In order to explain several anomalous features of some MCP stars,
and especially their extremely peculiar chemical composition (an
existence in atmosphere the short-lived radioactive elements) we
propose the following hypothesis. Let us consider an example of
HD101065 (Przybylski's star, PS), and assume that PS is a close
binary system with a non-seen companion being the neutron star
(NS). For this system an orbital plane is near perpendicular to
the line of sight (Fig. 1). The rapid wind, generated by NS, which
consists of the electron-positron plasma, is accelerated almost to
the speed of light and hit the PS atmosphere. The electron-positron
plasma falling on the PS must be ultrarelativistic, so that their
kinetic energy $E > m_{e}c^2$. The estimations show that the gamma-factor
of the "fast" electrons/positrons, which is necessary for starting the
photonuclear reactions, must exceed $20$. This estimation is
completely realistic, since gamma-factors of the electron-positron
plasma in the upper magnetosphere of a radio pulsar can be within
10-10000 (Ruderman \& Sutherland, 1975). In the spectrum of one of
the most studied pulsar PSR B0531+21 (Crab Nebula pulsar) there
really exist the gamma quanta within the required range of
energies (Fig. 2).

The high-energy electrons from electron-positron plasma can also
generate free neutrons via the direct interaction with hydrogen
nuclei in the PS atmosphere ($p+e^{-} \rightarrow n + \nu$). Such
free neutrons are necessary for the r-process to occur in the
considered medium.

Thus, the nuclei of heavy elements (including the radioactive
isotopes) in the PS atmosphere can be synthesized as a result of
two processes: the photonuclear reaction and neutron capture by
the seed nuclei of the lighter element. In both cases the source
of the energetic particles, that trigger  the nuclear reactions in
the PS atmosphere, is associated with the NS magnetosphere.

Let us show why the binary system PS-NS must be the close one. At
present there exist a large uncertainty in the mass estimations of
the PS. Thus, if we take one extreme estimate of its effective
temperature as 6600{$^\circ$} K, and consider another one
(spectral class B5, Perryman et al., 1997), then from the spectral
class-luminosity relation we obtain that mass of this star falls
in the range  $M_{PS}\approx 1.5 - 8\cdot M_{\odot}$. For the
further estimates we assume that masses of the PS and NS are
$M_{PS}=(1.5-8)M_{\odot}$ and $M_{NS}=(0.7-1.4)M_{\odot}$
respectively.

It is possible to estimate the parabolic velocity for this system
from the main integral of energy:
$V_{PAR}=\sqrt{\frac{2G(M_{PS}+M_{NS})}{R}}$ (were $R$ is the
distance from the center of mass to the NS). Taking into account
that the range of characteristic tangential velocities for the
known radio-pulsars is $V_{PSRs} \in \{100-500\}$ km/s, and the
fact that these velocities must not exceed the parabolic velocity
of the considered system, we obtain the range of the most probable
parameters (distance, the center of mass of the PS-NS system, the
orbital period, the corresponding masses of components). This
range is show in Fig. 3. It can be seen that the distances,
estimated in that way, ranges from 0.035 AU for a minimum mass
$M_{PS}$ to 1.0 AU (for maximum mass). Knowing the parallax of PS
$\pi\approx7.95\pm 1.07$ mas (Perryman et al., 1997), one can
estimate its distance $(D=125.8\pm15.7)$ pc and angular size (0.17
mas for minimum mass and 14 mas for maximum mass).

With parameters listed above, the orbital period of PS ranges from
261 days up to 0.7 days. Note that HR465, which has invisible
companion (but orbital plane is parallel to the line of sight) has
rotation period of 273 days (Scholz, 1978; Fig. 4). At the same
time,  the spectroscopic, photometrical data and magnetic field
measurements could be well represented with a period of 22-23 years
(Fuhrhmann, 1989).

Such a long-term variation can be caused by some active region on
the stellar surface that changes its position because of the weak
precession. In 1996 the magnetic field of this star was 5000 Gs,
and the lines of CrII were extremely strong. In 2004 the strong
lines of lanthanides and actinides (ThIII, UIII) were seen in the
visible part of spectra (Gopka et al., 2007). Magnetic field at
that time was 1300-1500 Gs (the estimate of Shavrina). Chromium
abundance changed by about 0.7 dex during the period of 8 years.
Within our model (MCP star + NS) active region (i.e. local
increase of the temperature, in particular) can be formed on the
surface of MCP stars as a result of a localized interaction of
atmosphere gas with relativistic plasma ejected by NS. To estimate
the local temperature increase one can use the value of kinematic
losses for known pulsars (Fig. 5) Resulting value is close to
3000{$^\circ$}- 6000{$^\circ$} K. The short-term light variation
(roAp phenomenon) can be also explained by such kind of
interactions.\\[2mm]

\begin{figure}
%\resizebox{4.26cm}{!}  % HD 101065 - NS orbit
\resizebox{\hsize}{!} {\includegraphics{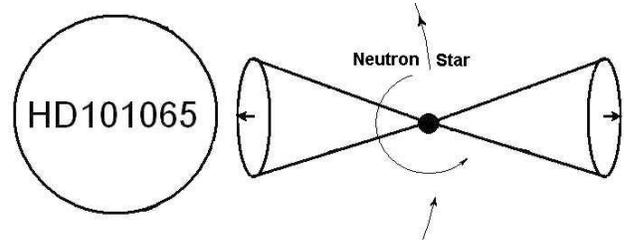}}
\label{hh} \caption{Geometry of the binary system containing Przybylsky's
star and neutron star.}
\end{figure}

\begin{figure}
%\resizebox{8.26cm}{!}
\resizebox{\hsize}{!} {\includegraphics{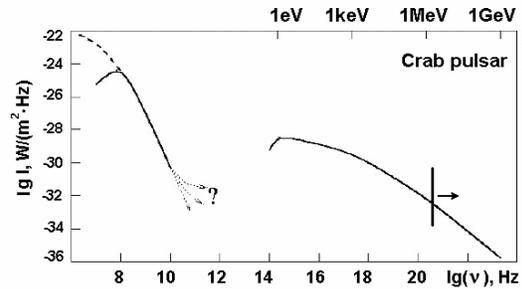}}
\label{hh} \caption{Spectrum of the Crab pulsar.}
\end{figure}

\begin{figure}
\resizebox{7.56cm}{!} {\includegraphics{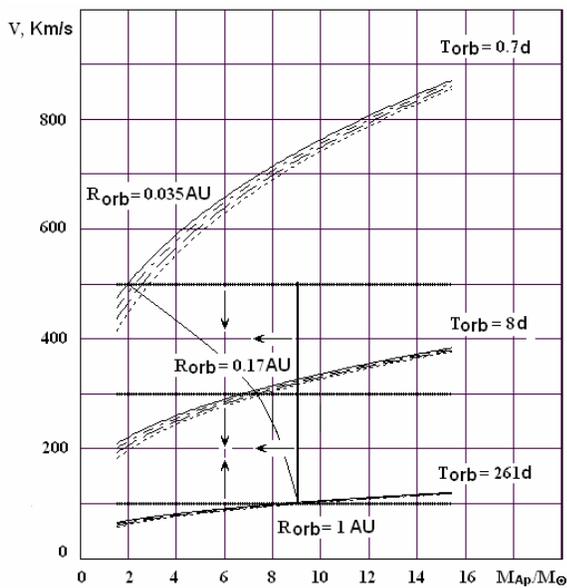}}
\label{hh}
%\resizebox{\hsize}{!} {\includegraphics{fig3-3.eps}} \label{hh}
\caption{Parameters of a close binary system Ap star - neutron star.}
\end{figure}

\begin{figure}
\resizebox{7.56cm}{!} {\includegraphics{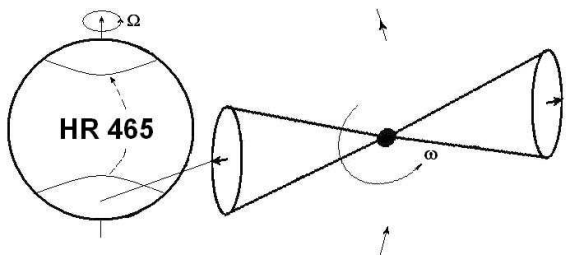}}
\label{hh}
%\resizebox{\hsize}{!} {\includegraphics{fig2-2.eps}} \label{hh}
\caption{Geometry of HR465 binary system. The position
of active area changes with period of  23 years.}
\end{figure}

\begin{figure}
\resizebox{6.26cm}{!}  {\includegraphics{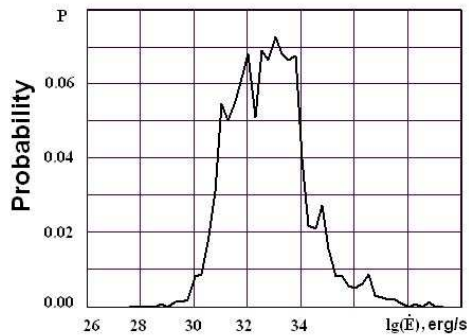}}
\label{hh}
%\resizebox{\hsize}{!} {\includegraphics{fig4-4.eps}} \label{hh}
\caption{The energy distribution of kinematic losses of 1627
pulsars.}
\end{figure}

{\bf 5. Conclusions}\\[1mm]

The proposed hypothesis, which consists in the supposition that
some MCP stars can be the binary stellar systems containing as a
secondary component neutron star, can be capable in natural
explaining of some peculiarities associated with these stars.
Among them: anomalous chemical composition, an existence of the
short-lived radioactive isotopes, short-time and long-period
variations of the light and magnetic field, X-ray and
radio-emission detected for some MCP stars.\\[1mm]

{\bf Acknowledgements}\\[1mm]

The authors  thank V.V. Ilyushin and V.V. Zakharenko, who helped
with useful discussion and remarks concerning this article.
Gopka V.F. was (partially) supported by research fund of
Ghonbuk  National University, Korea.\\[3mm]

\indent {\bf References\\[2mm]}

 Arnold M., Goriely S., and Takashi K.: 2007,  \h
 {\it Astrophysic},  {\bf5},  1.\\

 Babcock H.W.: 1949, {\it Observatory}, {\bf 69}, 191. \\

 Bagnulo S.: 2003, {\it IAU Symp. No 210.}, {\bf 210}, 9.  \\

 Bidelman A.: 2005,  {\it ASP Conf. series}, {\bf 336}, 309.  \\

 Burbidge E.M., Burbidge G.R., and Fowler  W.A.: \h
 1958,  {\it IAU Suppl.n6 \h
 Cambrudge Universyty press }, {\bf 6}, 222.\\

 Burbidge G.R.: 1965,  {\it Proc.of the IUA, \h
 Symp. n22}, 418 \\

 Burbidge  E.M., Burbidge G.R.: 1986, \h
 {\it Nucl. Astrophys.}, 22. \\

 Cowley C., Hatoog M.R. et al.: 1973, \h
 {\it Ap. J.}, {\bf183}, 127.\\

 Cowley C. \& Hubrig S.: 2005, \h
  {\it A. A.}, 2005, {\bf196},  21.\\

 Cowley C., Hubrig S., Castelli F.: 2005, \h
  {\it Contrib. Astron. Obs. \h
  Scalnete Pleso}, 2007, {\bf35}, 1.\\

 Cowley R.C., Bidelman W.P., et al.: 2004, \h
 {\it A. A.}, {\bf 419}, 1087.\\

 Drake S.A.: 1998,   {\it CoSka.},  {\bf27}, 382.\\

 Drake  S. A., Abbot D.C.,  et al.:  1987, \h
 {\it Ap. J.}, {\bf 322}, 902.  \\

 Drake S.A., Lynsky J.L., et al.: 1994, \h
  {\it Ap. J.},  {\bf420}, 387.\\

 Drake S.A., Linsky J.L., Wade G.A.: 2002, \h
 {\it AAS, Bull. of American Astron. Sosiety.}, \h
 {\bf34}, 1156.\\

 Fivet V., Quinet P., et al.: 2007 \h
  {\it MNRAS}, {\bf 380}, 771. \\

 Fuhrmann K.: 1989, {\it A. A. Suppl. Ser.}, {\bf80}, 399 \\

 Gopka V.F., Shavrina A.V.,  \h
 {Izv. KrAO}, {\bf104}, accepted\\

 Gopka V., Yushchenko A.,  et al.: 2005 \h
 {\it AIP Conf.Proc., Tokio, Japan}, {\bf 843}, 389. \\

 Gopka V., Yushchenko A.,  et al.: 2004, \h
 {\it IAU Symp. Poprad, Slovakia.}, {\bf  224}, 119. \\

 Goriely S.: 2007, {\it A. A.}, {\bf 466}, 619. \\

 Kurtz D.W. and  Martines P.: 2002, \h
 {\it  Baltic Astron. }, {\bf 9}, 253. \\

 Kurtz \& Wegner.: 1979, {\it Ap. J.}, {\bf 196}, 51. \\

 Landstreet  J.D.: 1990, {\it Ap. J.}, {\bf352}, 5. \\

 Leto P., Trigilio C., et al.: 2007, {\it A. A.}, {\bf 102},\h
 272.\\

 Lynsky J.L., Drake S.A., Bastian T.S.: 1989, \h
 {\it BAAS}, {\bf21}, 742.\\

 Lynsky J.L., Drake S.A.,  et al.: 1992 \h
  {\it Ap. J.}, {\bf393}, 341.  \\

 Mathys G.: 1991, {\it A.  A. Suppl. Ser. }, {\bf89}, 121  \\

 Michaud G.: 1970, {\it Ap.J.}, {\bf 160}, 641.  \\

 Perryman M.A.C.,  Lindegren L.,  et al.: 1997, \h
 {\it Astron. and Astrophys.},  {\bf 323}, \h L49.\\

 Richard O, Michaud G. \& Richard J.: 2002,   \h
 {\it Ap. J.}, {\bf 580}, 1100.  \\

 Ruderman M.A., Sutherland P.G. {\it Ap. J.}, \h
 1975, 196, N 1, 51 \\

 Scholz G.: 1978, {\it Astron. Nachr.},{\bf299}, 81 \\

 Shavrina A.V., Polosukhina N.S., et al.: 2003, \h
 {\it Astronomy Report},  {\bf 47}, 573. \\

 Shore \& Brown: 1990,  {\it Ap. J.}, {\bf 196}, 51.\\

 Stibbs D.W.N.: 1950, {\it MNRAS}, {\bf110}, 395. \\

 Trigilio C.: 2000,  {\it Ap. J.}, {\bf 196}, 51. \\

 Turcotte S., Richard O, Michaud G.: 1998, \h
 {\it Ap. J.}, {\bf 504}, 559.  \\

 Turcotte S., Richard O, et al.: 2000, \h
 {\it A. A.}, {\bf 272}, 559.  \\

 Quinet P., Argante C., et al.: 2007 \h
 {\it A. A.}, {\bf 474}, 307.  \\

 Wegner G., Petford A.D.: 1974, {\it MNRAS}, {\bf168}, 575. \\

\end{document}